\documentclass[final,5p,times, twocolumn]{elsarticle}

\usepackage{amssymb,amsmath,amsfonts,mathrsfs,bm,gensymb}
\usepackage{graphicx,float,subfig}
\usepackage{lineno}
\renewcommand{\figurename}{Fig.\,}
\renewcommand{\tablename}{Table\,}
\usepackage{geometry}
\usepackage{hyperref}
\usepackage{tabularx}
\usepackage{array}
\usepackage{color}
\usepackage{lipsum}
\usepackage{booktabs}
\usepackage[ruled, vlined]{algorithm2e}

\definecolor{gray}{RGB}{170, 170, 170}

\hypersetup{
    colorlinks=true,
    linkcolor=black,
    citecolor=black,
    filecolor=black,
    urlcolor=black,
}
\makeatother

\biboptions{sort&compress}

\newcommand{\norm}[1]{\left\lVert#1\right\rVert}
\newcommand{\pardydx}[2]{\frac{\partial #1}{\partial #2}}

\begin{document}

\begin{frontmatter}

\title{Shape-retaining beam-like morphing structures via localized snap through}

\author[]{Asifur Rahman}
\author[]{Samuele Ferracin\fnref{label2}}
\author[]{Sujata Tank}
\author[]{Chris Zhang}
\author[]{Paolo Celli\corref{cor1}}
\ead{paolo.celli@stonybrook.edu}

\address{Department of Civil Engineering, Stony Brook University, Stony Brook, NY 11794, USA}
% \address[add2]{Department of Mechanical Engineering and Applied Mechanics, University of Pennsylvania, Philadelphia, PA 19104, USA}
\cortext[cor1]{Corresponding author.}
\fntext[label2]{Current address: Department of Mechanical Engineering and Applied Mechanics, University of Pennsylvania, Philadelphia, PA 19104, USA.}

\begin{abstract}
In this study, we present a concept of morphing structure -- featuring an arch mounted on a compliant base -- that can be reconfigured via snap-through buckling and leverages bistability to retain its morphed shape. We show that one-dimensional arrays of such units yield beam-like structures that, upon localized snapping, can attain multiple, morphologically distinct stable shapes. Units are modeled using discrete elastic rods, a reduced-order formulation for beams and beam structures, and the results are validated via experiments. We leverage our model to understand the influence of the geometrical design parameters on the response and final shape of a unit. We then show that the morphed shapes of arrays of units can be predicted by concatenating results of simulations on single units, and leverage this idea to inverse-design structures that can be snapped into target stable shapes. Ultimately, our work suggests an up-scalable way to create shape-retaining morphing structures with target stable shapes.

\vspace{5px}
\normalsize{\textbf{This article may be downloaded for personal use only. Any other use requires prior permission of the author and Elsevier. This article appeared in}: \emph{International Journal of Solids and Structures} 300, 112917 (2024) \textbf{and may be found at}: \url{https://doi.org/10.1016/j.ijsolstr.2024.112917}}
\end{abstract}


\end{frontmatter}

% ======================================================================
\section{Introduction}
% ======================================================================

% GENERAL INTRO ON STRUCTURAL INSTABILITIES
Structural instabilities were traditionally regarded as a step towards failure; recent advances, however, demonstrate that they can be exploited to achieve functionality at various lengthscales \cite{reis2015perspective}. In this context, snap-through instabilities, whereby a structure can snap from an initial state to a second, potentially stable state, are particularly attractive and can be leveraged for energy absorption~\cite{restrepo2015phase, shan2015multistable, giri2021controlled}, vibration isolation~\cite{jiang2014snap, al2022advances} and shape morphing applications~\cite{rafsanjani2015snapping, haghpanah2016multistable, rafsanjani2016bistable}. In particular, systems capable of snap through that possess two or multiple stable states are termed bistable or multistable, respectively; their \emph{shape-retention} capacity is particularly useful in the context of deployable structures~\cite{pellegrino2001deployable}, and is essential to avoid the need for devices that keep structures in their stowed and deployed states.

Classical bistable systems owe their attributes to peculiar geometries that are conducive to snap-through mechanics, and to the presence of constraints \cite{vonMises1923}. For example, arch-like structures with boundary supports that fix the distance between the two end points are bistable \cite{rubin2004buckling}. Yet, the fact that the arch returns to its original shape if the constraints are removed is not ideal for deployable structures -- which need to be \emph{self standing} and not require supports to remain in their stable states. Self standing bistability can stem from a combination of geometry and material properties, or from geometry only. In the case of bistable composite plates and shells, constraints are imposed by the axial rigidity of the fibers, and by their particular orientation~\cite{iqbal2000bi}. Similar mechanisms are present in bistable systems featuring combinations of pre-stretched strips~\cite{chen2012nonlinear}, or stiff skeletons and pre-stretched strips \cite{risso2022highly, risso2021instability}. On the other hand, the self-standing nature of the stable states of bistable shells with folding lines~\cite{bende2015geometrically}, spherical caps~\cite{lapp2008exploring} and architected shells~\cite{chen2021computational, chen2021bistable} stems from geometrical attributes only.

% 'Fig.' label is defined by the elsevier documentclass
\begin{figure}[!t]
\centering
\includegraphics[scale=1.00]{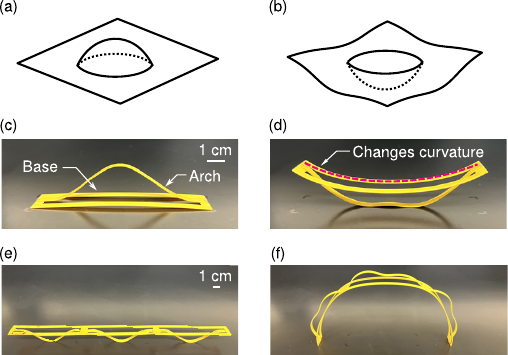}
\caption{Illustration of our idea and inspiration. Schematic of a bistable dimple on a soft sheet, in its (a) undeformed, and (b) snapped configurations. Fundamental bistable unit introduced in this work, in its (c) \textcolor{black}{stress-free first state, and in its} (d) second stable states. An array of three fundamental bistable units in its (e) undeformed, and (f) morphed configurations.}
\label{fig:inspiration}
\end{figure}

Among self-standing bistable systems, of interest to us is the dimpled sheet -- featuring spherical caps arranged on a thin plate according to regular patterns. While early incarnations of such system were made by impressing dimples onto metallic sheets~\cite{seffen2006mechanical}, recent versions are made of soft elastomers~\cite{faber2020dome, liu2023snap}. A single dimple, surrounded by a plate, is sketched in \figurename \ref{fig:inspiration}a. By pressing down on the dimple, this snaps and inverts its shape; this snapping is then accompanied by changes of curvature in the surrounding plate (\figurename \ref{fig:inspiration}b). When such units are arranged in two-dimensional (2D) patterns, the snapping of individual dimples affects the overall system, that tends to curve and form complex three-dimensional (3D) stable shapes~\cite{faber2020dome}. Interestingly, global shapes can be programmed by engineering the shape and spatial arrangement of dimples~\cite{udani2022taming}. Additionally, these systems display ``mechanical memory'', since different sequences of actuation yield entirely different morphed shapes~\cite{udani2022taming}. As a curiosity, dimpled sheets closely resemble popular PopIt fidget toys, where changes of sheet curvature are prevented by the presence of stiffening ribs. Although dimpled sheets display intriguing attributes, the high stresses involved in the snapping of a spherical cap make it challenging to translate these systems to aerospace or civil structural scales, or to actuate these structures via non-mechanical means (e.g., via temperature, humidity or piezoelectricity).

Here, we introduce a compliant structure inspired by dimpled sheets; our basic unit is shown in \figurename \ref{fig:inspiration}c,d in its first and second stable states, respectively. The unit features a sinusoidal arch bonded onto a rectangular, hollow base. By pushing down on the arch, the base provides enough lateral constraints to produce snap-through mechanics; in turn, snap through causes a change of curvature of the flexible base -- partially akin to what we observe in dimpled sheets. We develop a reduced order model for this unit, based on a discrete elastic rods (DER) formalism, validate the model via experiments, and use it to explore the parameter space of our structure -- thus illustrating its tailorable attributes. We then provide information on how to predict the shape-morphing capacity of one-dimensional (1D) linear arrays of cells by leveraging the model of a single cell, and show a pathway towards the creation of multistable, shape-retaining structures. Finally, we illustrate a simple inverse design procedure to produce 1D arrays that, when their units are locally snapped, morph into a target shape. Overall, we take a first step towards the creation of  structural analogs to bistable dimpled sheets by introducing a structure that, owing to its compliance, can be up-scaled.

Our article is organized as follows. In Section~\ref{sec:single}, we introduce the geometrical parameters of our unit, and study its mechanics via numerical simulations and experiments. In Section~\ref{sec:fk}, we postulate and test a conjecture that allows us to predict the deformed shape of a 1D array of units without explicitly modeling the entire array. In Section~\ref{sec:inverse}, we illustrate a framework to inverse design a 1D array to achieve morphing structures with target shapes. In Section~\ref{sec:extension}, we show preliminary evidence of up-scaling and introduce a 2D version of our unit. Finally, concluding remarks are given in Section~\ref{sec:conclusion}.

% ======================================================================
\section{Fabrication, design, and mechanics of a single unit}
\label{sec:single}
% ======================================================================

In this section we perform a complete analysis of the behavior of a single unit, via experiments and numerical simulations, and discuss details of the methods we employed. 

% ----------------------------------------------------------------------
\subsection{Fabrication process, parameters and properties}
\label{subsec:fabrication}

\begin{figure}[!htb]
\centering
\includegraphics[scale=1.00]{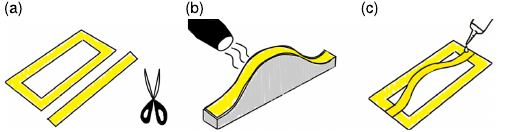}
\caption{Stages of the fabrication process. (a) Base and arch are cut from a PETG sheet. (b) The arch is taped to a sinusoidal mold with desired geometrical parameters and heat treated to preserve that shape after removal. (c) The arch and base are bonded together.} 
\label{fig:fabrication}
\end{figure}

The steps of our fabrication process for a single unit are illustrated in \figurename \ref{fig:fabrication}. The sinusoidal arch, which is initially a straight strip, and the rectangular base are laser cut from flat polyethylene terephthalate glycol (PETG) sheets as in \figurename \ref{fig:fabrication}a. 
The straight strip cannot be forced to form a sinusoidal arch by directly gluing it to the base, since this would force the base to have camber and would induce unwanted prestress in the system. To avoid any camber and prestress in our fabricated units, we heat treat the flat strip into a sinusoidal form. To do this, we 3D print a sinusoidal mold  with desired geometrical parameters and attach the flat strip to it via double-sided tape to take the shape of a sinusoid as shown in \figurename \ref{fig:fabrication}b. To have the strip retain its sinusoidal shape, we apply heat by using a heat gun. PETG sheets are very sensitive to this process; thus, heat should be applied with caution. The key is to apply heat uniformly on the strip from a distance of $\approx 1$ inch, and for multiple passes for a total of 2 minutes per strip. Applying heat on the same location for long times could cause unwanted warping or localized deformation. If the arch is under-heated and does not achieve its target shape, the base might camber when glued to the arch. The heating also deforms the molds, which are therefore single-use only. After the heating process is completed, we carefully remove the arch from the mold. Finally we bond the arch onto the flat base with cyanoacrylate glue as shown in \figurename \ref{fig:fabrication}c. If the heating process is done correctly, the arch retains its shape once removed from the mold and does not force the base to camber upon bonding.

\begin{figure}[!b]
\centering
\includegraphics[scale=1.00]{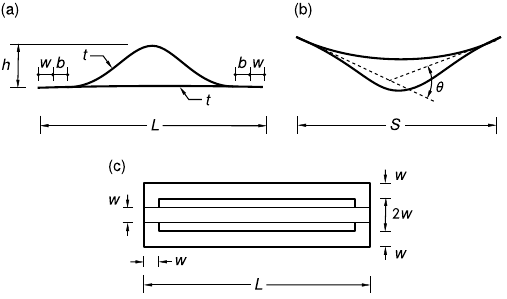}
\caption{The fundamental bistable unit, with its geometrical parameters. Side view of the unit in its (a) first (undeformed), and (b) second (deformed) stable states. (c) Top-down view of the undeformed configuration.}
\label{fig:single_unit}
\end{figure}

Schematic drawings of the orthographic views of our fundamental unit are shown in \figurename \ref{fig:single_unit}. 
In particular, \figurename \ref{fig:single_unit}a shows the side view of the unit in its unsnapped (undeformed, first stable state) configuration, while \figurename \ref{fig:single_unit}b shows the snapped configuration (second stable state), and \figurename \ref{fig:single_unit}c shows the top-down view of the unsnapped configuration. We fabricate specimens of base length $L = 100$ mm for three different arch heights $h =$ 10 mm, 20 mm and 40 mm. For each height, we construct specimens of thicknesses $t = 0.5$ mm and 0.75 mm (McMaster-Carr part numbers 9513K57 and 9513K59, respectively). The width $w$ of a strip in all specimens is set to 5 mm, which makes the overall width of the fundamental units to be 20 mm as shown in \figurename \ref{fig:single_unit}c. The arch is not fully sinusoidal throughout its length. Its extremities can be made flat for easier bonding to the base, and this flat region extends an amount $b+w$ on each side. We choose $b=w$ in all our specimens, unless otherwise noted.

Once we snap a unit to its second equilibrium state, the sinusoidal arch forces the straight base to adopt a curved shape. To quantify how much the base is deformed upon snapping we use two measures as shown in \figurename \ref{fig:single_unit}b. One parameter is the horizontal distance between the end points of the base, $S$, which is always smaller than $L$, and the other one is the angle $\theta$ between the tangents to the base at its endpoints. These parameters of the second stable shape are not measured from the specimens, but are computed from the numerical model during our parametric study and used to perform the inverse design. Finally, we set the following material properties for PETG:  Young's modulus $E = 2.2 \times 10^9$ Pa, density $\rho = 1.3\times 10^3$ kg/m$^3$ and Poisson's ratio $\nu = 0.38$. The Young's modulus is experimentally obtained by cutting beam like strips, bonding them to a shaker, subjecting them to vibrations at their root, by identifying the first resonant peak and by backtracking $E$ from a known formula of the first natural frequency of a cantilever beam.

% ----------------------------------------------------------------------
\subsection{Experimental setup}
\label{subsec:exp_setup}

\begin{figure}[!htb]
\centering
\includegraphics[scale=1.00]{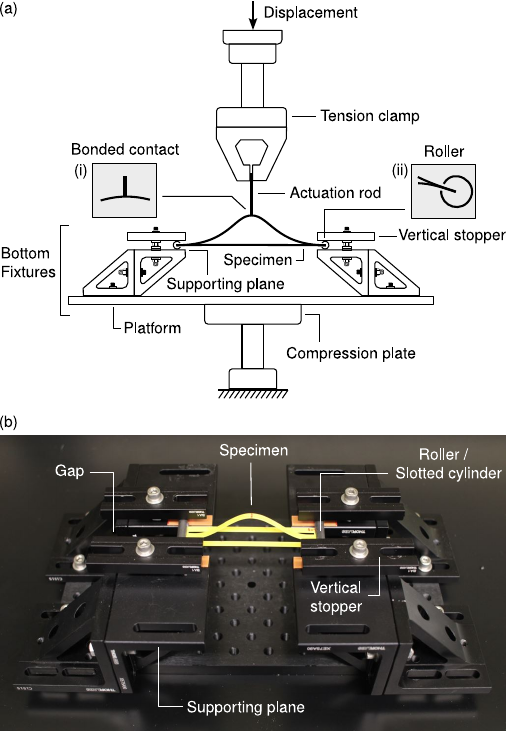}
\caption{Experimental setup for the mechanical tests on the behavior of a single unit. (a)  Schematic of the lateral view of the setup, where the details illustrate the custom fixtures and how the boundary conditions are implemented. (b) Photo (3/4 view) of the platform and fixtures. }
\label{fig:exp_setup}
\end{figure}

To capture the behavior of the fundamental units and to compare their responses to their numerical models, we perform displacement controlled compression tests on the fabricated specimens, by applying downward displacements at the apex of the sinusoidal arch and by restricting base motions. The maximum displacements applied go beyond the second equilibrium point to capture the entire snap-through force-deformation response. To perform the mechanical tests, we use an universal testing machine (UTM, Instron 68FM-100) with custom fixtures, and a schematic of the experimental setup is shown in \figurename \ref{fig:exp_setup}. 
The bottom fixture holds the structure and provides horizontal roller boundary conditions. It features four supporting planes, separated by gaps to provide room for a specimen to deform vertically beyond its second equilibrium. To allow for free lateral movement of the end points of the structure, we use two hollow slotted cylinders (McMaster-Carr part number 97161A199). We insert the edge of the units in these slits up to the center of the cylinder and use hot glue to keep them in position. We add vertical stoppers, that provide surfaces parallel to the supporting planes and prevent the cylinders from moving vertically.

In order to perform a true displacement-controlled test, we need to maintain contact with the specimens throughout the negative stiffness region. For this purpose, we use a T-shaped actuation rod laser cur out of $1/8$ in acrylic; the stem of the rod is bonded to the apex of the arch as shown in detail (i) of \figurename \ref{fig:exp_setup}a to provide a fixed boundary condition, and the web of the rod is gripped by the tensile clamp of the testing device. The length of the rod is chosen long enough to avoid contact between the tension clamp and the specimen, and its width is narrow enough to slip through the hole of the rectangular base of the specimens. The actuation rod is used to push the specimens downward at a rate of 1 mm/s, to implement a quasi-static loading scheme. 

The geometric parameters of the tested specimens are given in Table \ref{tab:parameters}. The values from the average height and average base length columns are used to create numerical models to compare to the experiments, in order to have the model account for some of the variability we have in our fabrication process.

\begin{table}[!htb]
\centering\small
\caption{Geometric parameters of the tested specimens for three different heights of 10 mm, 20 mm and 40 mm. For each height, we tested specimens of two different thicknesses (0.50 mm and 0.75 mm). {\color{black} The nominal base length for all specimens is 100 mm.}}
\begin{tabular}{p{1.25cm}p{1.25cm}p{1.35cm}p{1.25cm}p{1.25cm}}
    \toprule
     Nominal height (mm) & Thickness (mm) & Number of specimens tested & Average height (mm) & Average base length (mm) \\
     \midrule
     10 & 0.50 & 4 & 9.11 & 99.2 \\
        & 0.75 & 4 & 9.\color{black}{10} & 99.0 \\
     \midrule
     20 & 0.50 & 5 & 19.24 & 99.5 \\
        & 0.75 & 4 & 19.96 & 99.7 \\
     \midrule
     40 & 0.50 & 5 & 40.68 & 99.4 \\
        & 0.75 & 3 & 39.85 & 99.5 \\
     \toprule
\end{tabular}
\label{tab:parameters}
\end{table}

% ----------------------------------------------------------------------
\subsection{DER model}
\label{subsec:der_model}

To create a numerical model of the fundamental unit, we utilize Discrete Elastic Rods (DER), a discrete differential geometry-based modeling technique known for its  computational efficiency compared to the traditional finite element method (FEM). DER was first introduced in the computer graphics community \cite{bergou2008discrete} to simulate rod shaped objects such as hairs or brushes in real time, and has been successfully applied to structural and mechanics problems in a recent past \cite{lestringant2020modeling, huang2023bifurcations, baek2018form}. 

In choosing a DER-based model, we make the assumption that the mechanics of our 3D unit can be captured via a planar model, which we implicitly validate via comparison with experiments. Though the original DER technique can simulate a single rod, it is unable to model systems that have multiple interconnected rods. One of the simplest approaches to connect two rods is by inserting stiff springs between the connecting nodes \cite{qin2020genetic} -- which we adopt here. When adding stiff springs in the domain, it is important to choose their stiffness appropriately, and we do so by trial-and-error. The springs need to be stiff enough compared to the stiffness of the connecting rods, otherwise the connecting nodes would separate from each other; but they cannot be too stiff, which might lead to convergence issues. This method only allows to couple the translational degrees of freedom between the connected nodes. The coupling of twist between two rods is discussed in \cite{panetta2019x, perez2015design}, but it is not needed in this study due to the absence of significant twisting mechanics in our problem.

\begin{figure}[!htb]
\centering
\includegraphics[scale=1.00]{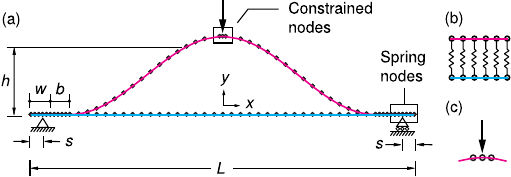}
\caption{(a) DER model of the fundamental bistable unit consisting of two rods (shown in red and blue). (b) The two rods are connected by stiff springs near the supports. {\color{black} (c) The load is applied at the center node of the arch, the adjacent two nodes on both sides are constrained to move together in the vertical direction.}} 
\label{fig:discretization}
\end{figure}

The schematic of the model of our fundamental unit, which consists of two rods (one for the sinusoid and one for the flat base) is shown in \figurename \ref{fig:discretization}a. 
The two rods are coupled by six stiff springs at each end, as illustrated in \figurename \ref{fig:discretization}b. The spring nodes from the two rods are almost at the same location, separated by a small distance of $10^{-6}$ m to avoid numerical singularities. The rods have rectangular cross section of height equal to the thickness of the PETG sheet (either 0.50 mm or 0.75 mm depending on the specimen). For the sinusoid, which consists of a single strip, we use $w$ as cross sectional width for its entire length. Modeling the hollow rectangular base using a single rod requires variable width to capture its stiffness distribution. Thus, for the base rod, we set a width of $4w$ in the non-hollow regions (near the supports), and $2w$ everywhere else (see \figurename \ref{fig:single_unit} for reference). By modeling the base this way, we assume little-to-no variability in its behavior in the width direction. The boundaries are applied at 3 mm offset from the ends to represent the behavior of the rollers in our experimental setup. A downward vertical load is applied on the central node of the arch. The central three nodes of the sinusoid are modeled as a rigid joint by coupling their degrees of freedom, {\color{black} as shown in \figurename \ref{fig:discretization}c,}  to represent the bonded contact as shown in \figurename \ref{fig:exp_setup}. Additional details on our DER implementation, and some basics on DER for beginners, are reported in \ref{app:der}.

% ----------------------------------------------------------------------
\subsection{Behavior of a single unit, and model limitations}
\label{subsec:results}

\begin{figure*}[!htb]
\centering
\includegraphics[scale=1.00]{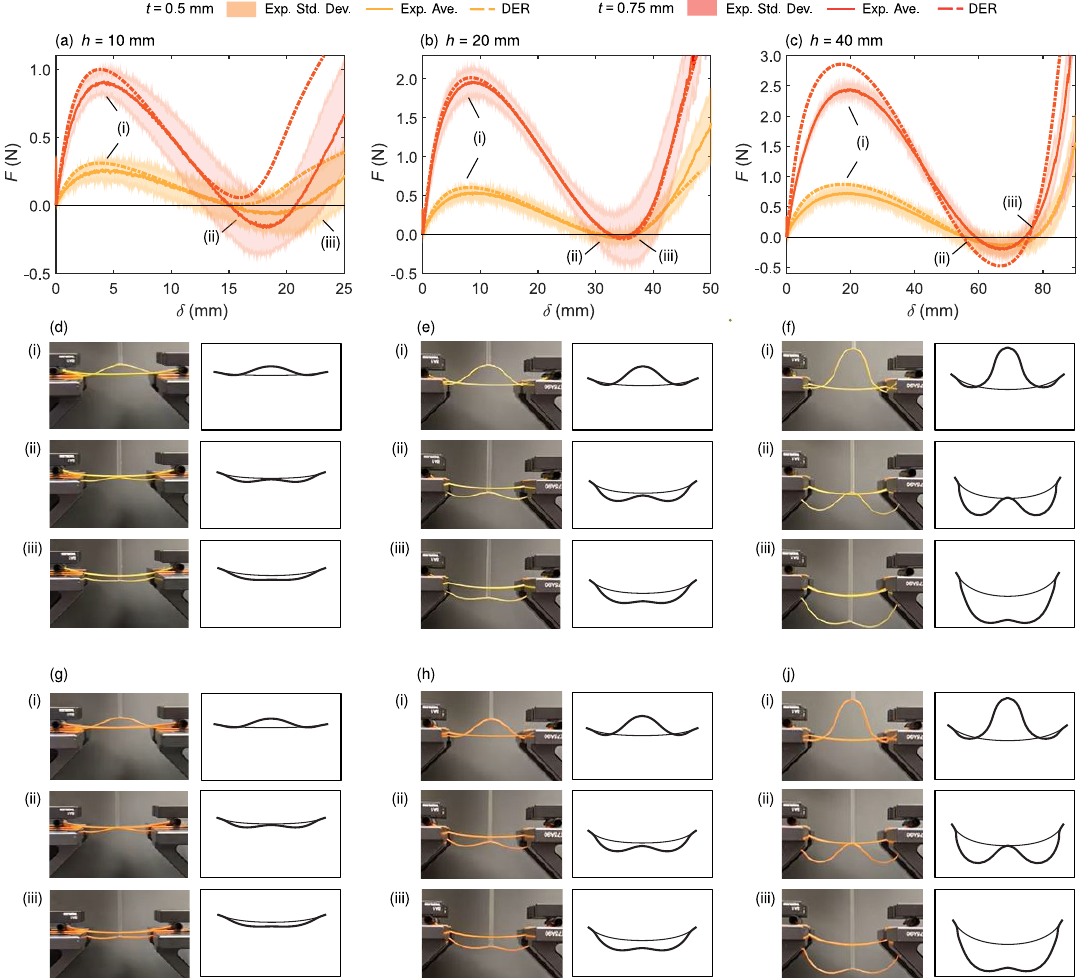}
\caption{Comparison between numerical and experimental results for a single unit. (a)-(c) Force-deformation curves for all specimens, arranged by arch height: (a) $h=10$\,mm, (b) $h=20$\,mm, (c) $h=40$\,mm. For the experimental curves, the continuous lines represent averages of all samples of the same type, and the shaded areas are the standard deviations. Individual force-deformation curves for all specimens are reported in \figurename \ref{fig:forcedef_raw}. (d)-(j) Numerical and experimental deformed shapes at the deformation levels corresponding to (i) the critical force, (ii) the unstable equilibrium, and (iii) the second stable equilibrium. Deformed shapes are organized in a table format, with each column corresponding to a different value of $h$, and each row corresponding to a different thickness $t$, such that: (d) $h=10$\,mm, $t=0.5$\,mm; (e) $h=20$\,mm, $t=0.5$\,mm; (f) $h=40$\,mm, $t=0.5$\,mm; (g) $h=10$\,mm, $t=0.75$\,mm; (h) $h=20$\,mm, $t=0.75$\,mm; (j) $h=40$\,mm, $t=0.75$\,mm.} 
\label{fig:forcedef}
\end{figure*}
\begin{figure*}[!htb]
\centering
\includegraphics[scale=1.00]{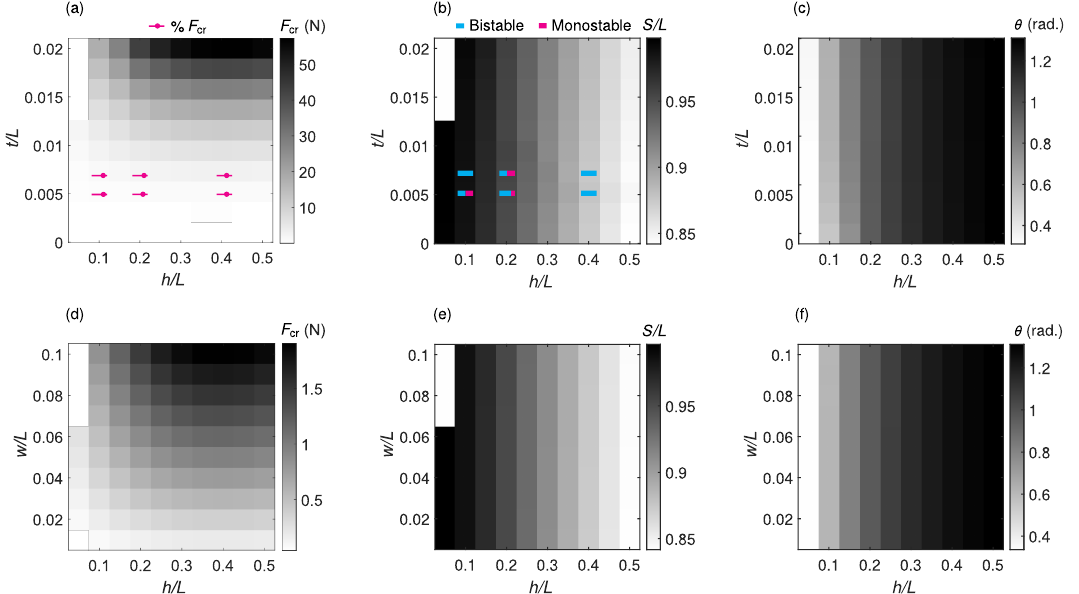}
\caption{Parametric maps of normalized height ($h/L$) vs.\ normalized thickness ($t/L$), with fixed $w/L=0.05$, for (a) critical force $F_\text{cr}$, (b) normalized distance between base end points at the second equilibrium $S/L$, (c) angle between end tangents at the second equilibrium $\theta$. For the parameter combinations tested in \figurename\ref{fig:forcedef}, the slider markers in (a) indicate how well the experiments compare to the numerics and the two-color bars in (b) indicate how many fabricated specimens are bistable. These markers can help interpret the numerical results. Parametric maps of normalized height ($h/L$) vs. normalized width ($w/L$), with fixed $t/L=0.005$, for (d) critical force $F_\text{cr}$, (e) normalized distance between base end points at the second equilibrium $S/L$, (f) angle between end tangents at the second equilibrium $\theta$.} 
\label{fig:maps}
\end{figure*}

Numerical and experimental results for a single unit and for various combinations of geometrical parameters (sheet thickness $t$ and arch height $h$, all shown in \figurename \ref{fig:single_unit}) are compared in \figurename \ref{fig:forcedef}. All force-deformation curves are shown in \figurename \ref{fig:forcedef}a-c, while the numerical and experimental deformed shapes at a few key locations are shown in \figurename \ref{fig:forcedef}d-j. Deformed shapes marked (i) correspond to the critical force location, i.e., to the point of maximum force before snap through; the ones marked (ii) correspond to the unstable equilibrium location, where the curve crosses the zero force axis with a negative slope; the ones marked (iii), instead, correspond to the second stable equilibrium, where the curve crosses again the zero-force axis with a positive slope. In general, numerical force-deformation curves show larger force than the experimental ones at the same deformation level; the degree of agreement between the two sets of curves varies depending on the arch height, with $h=20$\,mm being the case that matches better. At a first glance, numerical and experimental deformed shapes agree very well across arch heights and thicknesses. \textcolor{black}{The overall symmetry of the deformed shapes, which is atypical in true single-point indentations experiments, is due to the actuating rod being glued to a finite portion of the arch; yet, some experimental images show loss of symmetry, which is due to} fabrication imperfections and positioning errors of the specimen within the setup. This loss of symmetry, visible across specimen heights and thicknesses and typically associated with higher buckling modes, provides a reasonable explanation for the lower forces recorded in experiments~\cite{vangbo1998analytical}. More details on these results are given below.

Looking at the experimental results for a 10\,mm arch height in \figurename \ref{fig:forcedef}a, we observe that the standard deviation relative to the force values involved is larger compared to the other heights, which indicates that it is harder to consistently fabricate specimens with these characteristics. This aspect is further confirmed by looking at the experimental force-deformation curves for individual specimens, shown in \figurename \ref{fig:forcedef_raw}a,b of \ref{app:forcedef}. From these individual curves, we can see that two out of four specimens with 0.50 mm thickness are bistable, and all four 0.75 mm thickness specimens are bistable. Moreover, specimens at this arch height present large variations in response both in terms of critical force value and in terms of location of the equilibrium points (points where the curve crosses the zero force axis) -- which explains the large standard deviations in \figurename \ref{fig:forcedef}a. The numerical responses for the same height, shown in \figurename \ref{fig:forcedef}a, do not capture well the experimental behavior after the critical force, since the curves fall outside the standard deviation for both thicknesses, to the point that they fail to predict the occurrence of bistability (by crossing the zero force axis only at the origin). While our model is computationally efficient, its reduced-order nature does not allow us to elaborate further on these results, and to pin down the reason behind the discrepancy. We simply take note of this fact, and of the fabrication difficulties, to limit the range of arch heights we will consider in the remainder of this work. 

Moving to the intermediate height of 20 mm, corresponding to the force-deformation curves in \figurename \ref{fig:forcedef}b, we have great match in the entire deformation domain for both thicknesses, with the numerical response always falling within the bounds of the experimental standard deviation. Thus, the critical forces and the locations of the unstable and second stable equilibria all match well. By looking at the individual force-deformation curves in \figurename \ref{fig:forcedef_raw}c,d of ~\ref{app:forcedef}, we can see that the specimens have much more uniform behavior with respect to the $h=10$\,mm case. However, we can still see that fabricating specimens that are consistently bistable is challenging: three out of four specimens with 0.50 mm thickness are bistable, and two out of four of the 0.75 mm thick ones are bistable.

Finally, the force-deformation curves for an arch height of 40 mm are shown in \figurename \ref{fig:forcedef}c. Looking at the experimental curves, we can see that they have the lowest relative standard deviation; this aspect is even clearer by looking at the individual curves in \figurename \ref{fig:forcedef_raw}e,f of ~\ref{app:forcedef}, where the consistency between curves is clear. This aspect is further remarked by the fact that all specimens for this height are bistable. These observations indicate that fabrication inconsistencies, which are bound to occur for all specimen types, play a lesser role as we increase the arch height and as the forces involved during the snap-through process increase in magnitude. The numerical curves capture well the location of the crossings with the zero force axis, while over-predicting force values by larger amounts with respect to the other arch heights. We ascribe this increased discrepancy to two aspects. On one hand, despite their remarkable consistency, samples for this arch height are also consistently more asymmetric, with the arch visibly tilted in the width-direction; in turn, this could be responsible for a stronger knockdown of the experimental curves. On the other hand, as the arch gets higher, the base gets more stressed during snap through. In turn, this could cause localized deformations at the base, in the neighborhood of the arch bonding sites -- an aspect that our simplified numerical model cannot capture and that would lower the force-deformation curve.

Looking at these results holistically, we can draw a few concluding remarks, which will be useful when interpreting the parametric study of Section~\ref{subsec:parametric}: 1) we observe more consistent bistability as we increase arch height, despite the fact that these specimens feature larger imperfections; 2) the model does better in predicting experimental force values for intermediate arch heights; 3) the model does well in terms of predicting geometric features of the deformed sample (location of second equilibrium) beyond a lower bound located between 10 and 20 mm. In light of these observations, the arch height to be chosen strongly depends on the application we are interested in. If our intention is to use our unit to create a switch or another device that snaps at the expected force, then an intermediate arch height would be the best choice. If we are instead interested in using these units to create shape-morphing structures, as we are in this article, then we should select arches higher than 10 mm. Furthermore, if we are only interested in creating structures that morph into target shapes, then the force inconsistencies we have at high $h$ do not matter. Instead, if we are interested in creating morphing structures with predictable response to external forces in their second state, then the high-$h$ inconsistencies matter.

% ----------------------------------------------------------------------
\subsection{Parametric study}
\label{subsec:parametric}

After building some confidence in our model, we are ready to study how different design parameters affect the response of our structural unit. From all variables shown in \figurename\ref{fig:single_unit}, we decide to investigate the effect of the independent parameters height $h$, thickness $t$, and width $w$ on the dependent parameters critical force $F_\text{cr}$, span $S$ and angle between the end tangents $\theta$ at the second equilibrium point. As we do not have any analytical model of our structure, we perform our parametric study by running the numerical analysis multiple times. The results of this study are shown in \figurename\ref{fig:maps}.
In this study, we set $b=0$.
%The heights, thicknesses and widths for this study range from $L/10$ to $L/2$, $L/20$ to $L/5$ and $L/50$ to $L/10$ respectively. 

\figurename \ref{fig:maps}a shows the critical forces $F_\text{cr}$ for different $h$ and $t$ combinations. In this plot, we set $w/L = 0.05$. We see that $F_\text{cr}$ -- which we do not normalize to keep the trends as clean as possible -- is a function of both $h$ and $t$, and it increases as $h$ and $t$ both increase. This makes sense since increasing both parameters causes the structure to be stiffer. We overlay slider markers to the colormap to point out how well the experiments match the numerics for the parameters listed in Table \ref{tab:parameters} and corresponding to the results of \figurename\ref{fig:forcedef}. The sliders in the bottom row represent thickness 0.50 mm and those in the top row correspond to 0.75 mm. These markers highlight that we should be more confident about predicted $F_\text{cr}$ values for the left half of the parametric map. \figurename \ref{fig:maps}b shows the span length at the second equilibrium $S$, normalized by $L$, and \figurename \ref{fig:maps}c shows the angle between the end tangents at the second equilibrium $\theta$ for the same set of $t$, $h$ and $w/L$ values. Unlike critical forces, which are mechanical properties of the system, $S$ and $\theta$ are kinematic properties which only depend on the second stable shape. As expected, they are only affected by $h$, while remaining approximately invariant with respect to $t$. In particular, as $h$ increases, $S$ decreases as shown in \figurename \ref{fig:maps}b because taller sinusoidal arches force the base to contract more to reach the second equilibrium. Similarly, as $h$ increases, the sinusoidal arch also forces the base to curve more, as shown by the increase of angle $\theta$ in \figurename \ref{fig:maps}c. We overlay two-color bars on \figurename \ref{fig:maps}b to represent how many of the fabricated units are bistable for the sets of parameters of Table \ref{tab:parameters} and corresponding to the results in \figurename\ref{fig:forcedef_raw}. We can see that, as the height increases, the units are more consistently bistable for both thicknesses. For lower heights, the map shows monostable regions which should be avoided when designing bistable structures.

\figurename \ref{fig:maps}d-f are analogous maps to \ref{fig:maps}a-c, with the only difference being that we analyze the effect of width $w$ instead of thickness $t$. In these plots, we set $t/L = 0.005$. \figurename \ref{fig:maps}d shows that $F_\text{cr}$ depends on $w$ as expected because beams and arches of larger width have a larger bending stiffness. We can then see that \figurename \ref{fig:maps}e,f display trends that are similar to \figurename \ref{fig:maps}b,c; thus, just like the thickness $t$, the width $w$ does not play a key role in determining the kinematic properties $S$ and $\theta$. We will leverage the indifference of kinematic properties on $t$ and $w$ when performing inverse design of structures with target shapes. During the inverse design process, we will constantly refer to these design maps to understand which units to use to obtain desired local curvatures.

% ======================================================================
\section{Forward kinematics of arrays}
\label{sec:fk}
% ======================================================================

After exploring the mechanics of a single unit, we now focus on the response of 1D arrays of units. Arrays are constructed by concatenating multiple units, identical or with different dimensions, one after another. Since each unit can be snapped independently of the others, the array displays many different stable states depending on which units are snapped -- an aspect that makes them intriguing as multistable shape-morphing structures. Since an array has multiple units, modeling it would result in higher computational cost than a single unit. Additionally, choosing boundary conditions to be able to sequentially snap units in the array is far from trivial. 

\begin{figure}[!htb]
\centering
\includegraphics[scale=1.00]{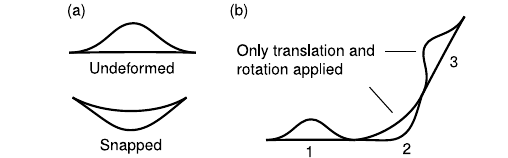}
\caption{The process of forward kinematics. (a) The two stable configurations of the fundamental unit: undeformed and snapped. (b) Only translations and rotations of the stable configurations are required to obtain the deformed shape of an array of fundamental units with any arbitrary snapping sequence.}
\label{fig:fk_steps}
\end{figure}

In response to this challenge, we postulate a conjecture that the deformed shape of an array of fundamental units, undeformed or snapped in any arbitrary sequence, can be simply obtained by concatenating the undeformed and deformed shapes of a single unit. This conjecture, which is illustrated in \figurename \ref{fig:fk_steps}, would allow to predict the final shape of an array without explicitly modeling it, thus saving computational cost. 
This conjecture is based on the observation that neighboring units don't interact, and is not valid in 2D systems like dimpled sheets, where snapping a unit strongly affects its neighbors. \figurename \ref{fig:fk_steps}b shows an array of three units where only the second one is snapped. According to the conjecture, we can predict the shape of this array by translating, rotating and concatenating the basic shapes. First, we place the undeformed shape (from \figurename \ref{fig:fk_steps}a) that corresponds to unit 1. Second, we translate the snapped shape (from \figurename \ref{fig:fk_steps}a) and match the end point of unit 1, then rotate to match the tangents at the contact point to obtain unit 2. Third, to obtain unit 3, we take another undeformed unit, translate it to match the endpoint of unit 2 and rotate it to match the tangents at the contact point. 

\begin{figure*}[!htb]
\centering
\includegraphics[scale=1.00]{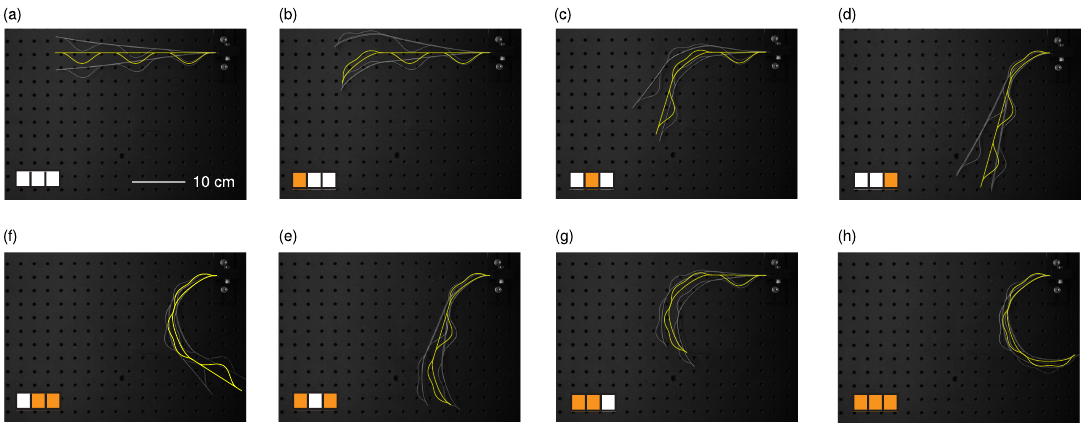}
\caption{Experimental validation of forward kinematics conjecture. An array consisting of three units, having a total of $2^3 = 8$ stable states is shown in all these states from (a) to (h){\color{black}, and the corresponding snapped units are indicated by the inset in each figure}. The predictions of the numerical model, shown in yellow, are overlaid on the photographs of the actual specimens, shown in white. As friction plays a role in the experiment, we show photos of two extreme positions of the array, obtained by pushing the array in one direction and then the other.}
\label{fig:fk}
\end{figure*}

The experimental validation of our conjecture for an array composed of three identical units (with parameters $t=0.5\,$mm, $h=20\,$mm, $w=5\,$mm, $L=100\,$mm) is shown in \figurename \ref{fig:fk}. 
As we do not implement gravity in our model, we cannot place the array in upright position since gravity would affect its deformed shapes. Therefore, we place the array sideways and clamp one end of it, eliminating deformations due to gravity. However, some inevitable friction between the specimen and the underlying surface is present in this setup. Since each unit can be in its undeformed or snapped state, there are eight possible permutations of this structure, from \figurename \ref{fig:fk}a  where no unit is snapped to \figurename\ref{fig:fk}h where all units are snapped. The numerically predicted shapes, obtained via our conjecture, are overlaid on top of the photographs of the physical specimens, for each configuration. Due to the friction between the array and the table surface, the deformed shape has multiple achievable positions. For each configuration, we took two photos shown as overlay in \figurename \ref{fig:fk}, which represent two extreme positions of the array due to friction. The positions were obtained by gently pushing the array in one direction and then in the other direction. From the comparisons, we can see that the numerical prediction falls withing the positional bounds imposed by friction for all snapping configurations. We consider this as a successful validation of our conjecture, which will be heavily leveraged for inverse-design.

% ======================================================================
\section{Geometrical inverse design}
\label{sec:inverse}
% ======================================================================

\begin{figure}[!htb]
\centering
\includegraphics[scale=1.00]{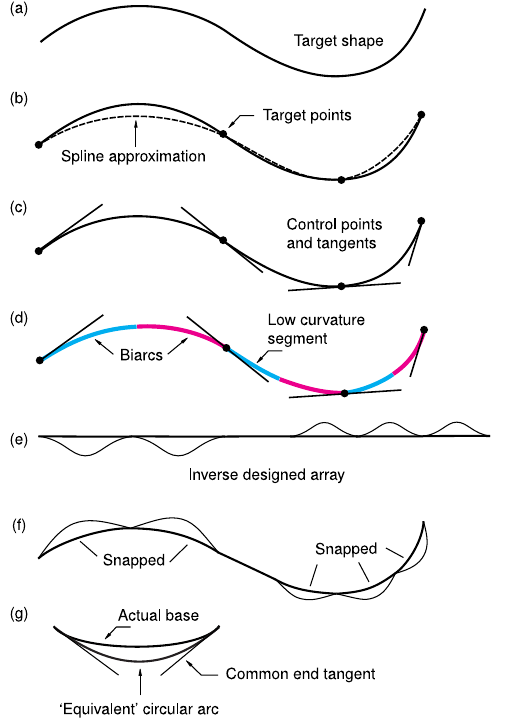}
\caption{Geometric inverse design process. (a) We begin with a target shape, that needs to be matched by the deformed base of an array of snapped units. (b) To determine the size, parameters and the number of units, we select some arbitrary target points on the target shape, and construct a spline through those points, which will give us tangents as in (c). If the target shape is simple and the tangents at the target points are known, our procedure starts at (c). (d) Knowing the spline tangents at some control points, we construct arcs of circles (biarcs) that match those tangents. Segments with small curvature, corresponding to difficult-to-fabricate units with shallow arches, are replaced with straights lines. From the end-to-end lengths and tangent angles of the biarcs, we resort to the parametric maps of \figurename\ref{fig:maps} to determine the parameters of each unit of the array. A sample inverse-designed array is shown in (e) in its undeformed state and in (f) in its deformed state -- in which the base matches the target shape in (a). (g) Even though the shape of the base of one of our units in its snapped state is not perfectly circular, we use arcs of circles in our inverse design process since we can easily identify an equivalent circular arc corresponding to an actual base.  }
\label{fig:inverse_steps}
\end{figure}

We now explore the idea of inverse design of an array of fundamental units so that, when these are snapped, the structure matches a target shape. In general, inverse design procedures are important to unleash the full programmability potential of morphing structures, and to make them attractive for applications~\cite{konakovic2018rapid, jin2020kirigami, zhang2022shape, dang2022inverse}. For a selected target shape, we need to determine how many of these units are required to match that shape, and their dimensions. As we can see from \figurename\ref{fig:maps}e,f, the width $w$ and thickness $t$ of our units do not play any role in the kinematic properties; we can exploit this property by keeping these parameters constant in our arrays, while varying $h$ and $L$ from unit to unit. The procedure we follow for our inverse design, partially inspired by works on inverse kinematics for robotics~\cite{webster2010design} and on morphing structures~\cite{liu2019encoding}, is summarized in \figurename \ref{fig:inverse_steps}.

We begin with a target shape, shown in \figurename \ref{fig:inverse_steps}a, that needs to be matched by the base of a 1D array of units once all of its arches are snapped. To identify tangents to this curve, which are needed to identify what units we should use, we choose some target points on the shape and determine the spline passing through these points, as shown in \figurename \ref{fig:inverse_steps}b. Then, by using the spline equation, we determine the tangents at desired control points (in this figure, control points coincide with target points, but this need not be the case). Alternatively, if we already know the tangents at some points of a target shape, we can directly start from \figurename \ref{fig:inverse_steps}c, which eliminates the need for a spline approximation. Please note that, when we use the spline interpolation, our algorithm is limited and can only approximate relatively-shallow shapes without overlaps along the vertical direction; additionally, the process of identifying control points is somewhat arbitrary, and we do it manually.

The control points and their tangents are then used to construct biarcs as illustrated in Fig \ref{fig:inverse_steps}d. A biarc is a smooth curve created from arcs of constant curvature, which are essentially circular curves with matching tangents at their connecting points. After constructing the biarcs, we can determine the angles between the tangents at the end points; knowing these angles and the distance between points (which correspond to parameters $\theta$ and $S$ in our units, as shown in \figurename \ref{fig:single_unit}), we can use the parametric maps of \figurename\ref{fig:maps} to select the parameters of each unit to match that deformed shape. In particular, we use our $\theta$ requirement to enter the map in \figurename\ref{fig:maps}f (at any $w/L$, due to the non-influential nature of this parameter) and determine the desired $h/L$; by using this $h/L$, and the $S$ requirement, we can then determine $L$ from \figurename\ref{fig:maps}e.

The approach of constructing biarcs may result in very small curvature for a segment; from our studies on a single unit, we know that small curvatures correspond to units with shallow arches, which are difficult to fabricate and not consistently bistable. When this happens, our algorithm replaces a low-curvature segment with a straight line, which can be obtained by using an arch-less unit. \figurename \ref{fig:inverse_steps}e shows a sample of inverse-designed array in its undeformed configuration, and \figurename \ref{fig:inverse_steps}f shows the same array in its deformed configuration, that matches the required shape. As shown in Fig \ref{fig:inverse_steps}g, it is important to note that when a unit is snapped, the base is not a circular segment. However, knowing the tangent angles and the distance between end points allows to identify an equivalent circular arch -- which justifies the biarc approximation of the target shape.

\begin{figure*}[!htb]
\centering
\includegraphics[scale=1.00]{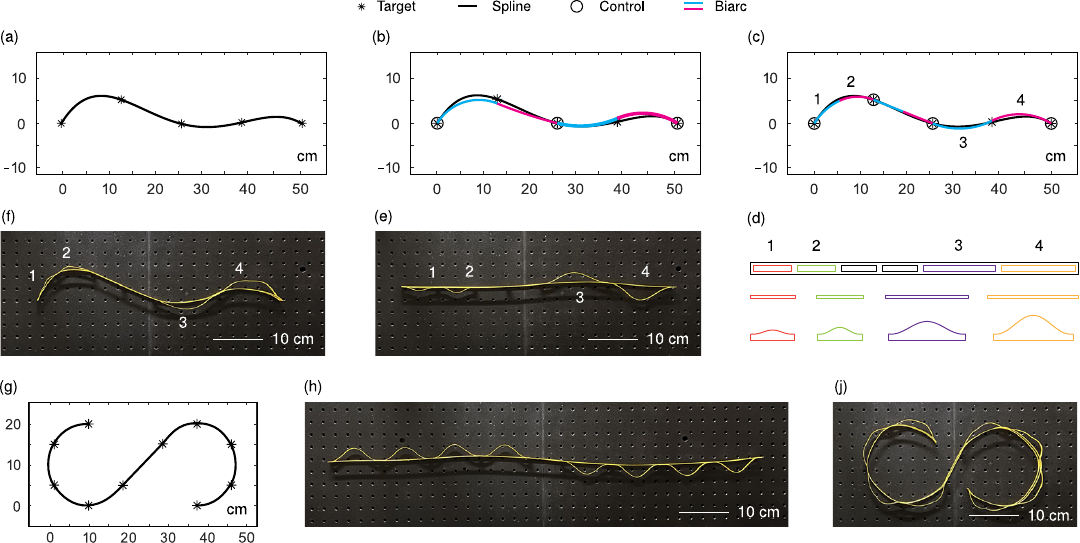}
\caption{Geometric inverse design of two target shapes. For the asymmetric M-like shape, we illustrate the entire inverse design process from (a) to (f); for the S shape, we only show the target shape and end result from (g) to (j). (a) We define the asymmetric M shape by some arbitrary points, and construct a spline though these input points. From the spline equation, we determine the tangents at some control points, which are used to construct biarcs. (b) Shows the first iteration of the biarc interpolation. (c) Second iteration of the biarc interpolation, where we update the design by replacing low-curvature arcs with straight segments; the final biarcs are labeled as 1 to 4. (d) Base strip, arch strips and outlines of the molds to form the arches, for each unit, obtained by using the parametric maps showed in \figurename\ref{fig:maps}. The fabricated structure for the M shape, in its (e) undeformed state, and (f) snapped state. (g) Outline of the S shape, for which we directly input the control points and their corresponding tangents. The fabricated array for the first target shape is shown in its (h) undeformed state, and (j) deformed state.}
\label{fig:inverse_experiment}
\end{figure*}

\figurename\ref{fig:inverse_experiment} shows the experimental validation of the inverse design process. We validate the approach both with and without spline approximation. We select two shapes for the validation; one is a shallow, asymmetric M-like shape, to be approximated with splines, and the other is a symmetric S shape, which is regular enough not to require splines. For the M shape, we start from five equally spaced target points and construct a spline approximation of the shape as shown in \figurename\ref{fig:inverse_experiment}a. Then, we select three control points along the spline (at the beginning, center and at the end); using a bi-arc interpolation function on these points and their tangents results in four biarcs, as shown in \figurename\ref{fig:inverse_experiment}b. The second arc from the left has very low curvature, which would result in a hard-to-fabricate unit, and is therefore replaced with straight segments. We do this automatically by adding another control point between the left-most and the center one, and refine our design. \figurename\ref{fig:inverse_experiment}c shows the second iteration of the design process, which results in four biarcs labeled as 1 to 4 and two straight segment between 2 and 3. After constructing the biarcs, we can determine the distance between end points $S$ and the angle $\theta$ between the tangents to their end points -- which are used to determine the height and length of each unit from \figurename\ref{fig:maps}. Based on the results of this search of our design space, our algorithm outputs laser cut drawings for the base strip, arch strips, and the mold drawings to be extruded and 3D printed, as shown in \figurename\ref{fig:inverse_experiment}d. The actual dimensions of these units are listed in \tablename\ref{tab:inverse_design}. \figurename\ref{fig:inverse_experiment}e shows the fabricated and assembled array for the M shape. Note that the arches for units 3 and 4 are on opposite sides of the base to obtain the desired shape once the units are snapped. The deformed shape of the snapped array is shown in \figurename\ref{fig:inverse_experiment}f. By visually comparing the shape of the deformed base in \figurename\ref{fig:inverse_experiment}f with the target shape in \figurename\ref{fig:inverse_experiment}a and with the expected shape of \figurename\ref{fig:inverse_experiment}c, we can conclude that the match is satisfactory.
\begin{table}[!htb]
\centering\small
\caption{Geometric parameters for the inverse-designed units shown in \figurename\ref{fig:inverse_experiment}.}
\begin{tabular}{p{1.00cm}p{1.00cm}p{1.5cm}p{1.25cm}}
    \toprule
     Shape & Unit & Base length (cm) & Height (cm) \\
     \midrule
     M & 1 & 7.26 & 0.69 \\
       & 2 & 7.38 & 1.13 \\
       %& * & 6.73 & N/A
       & 3 & 12.94 & 2.11 \\
       & 4 & 13.32 & 3.13 \\
     \toprule
     S & All & 10.67 & 2.58 \\
     \toprule
\end{tabular}
\label{tab:inverse_design}
\end{table}
For the S shape, we only show the input and the end result. The target shape is defined by four circular segments on each side connected by a straight segment in between. As these are circular segments, we can compute the tangents at the target points -- which eliminates the need for a spline approximation for this shape. We then follow the same procedure used for the M shape and obtain the size and parameters of all the units, together with drawings for laser cutting and 3D printing. Since the circular segments are all equal in size, all units feature the same set of parameters listed in \tablename\ref{tab:inverse_design}. The fabricated array is shown in its undeformed state in \figurename\ref{fig:inverse_experiment}h and in its deformed state in \figurename\ref{fig:inverse_experiment}j. Please note that, once again, the arches for one arm of the S shape are mounted on one side of the base, while the arches for the other arm are mounted on the opposite side, in order to obtain curvatures in opposite directions. \textcolor{black}{Additionally, since the base of the entire array is too long for our laser cutter, we cut two halves of the base and glue them together to get the structure in \figurename\ref{fig:inverse_experiment}h}. While the deformed specimen indeed resembles an S shape, the match is not entirely satisfactory since the arms are more curved than expected. This curvature overshooting is likely obtained by compounding acceptable errors in consecutive units -- something that can only be overcome by reducing fabrication and modeling inaccuracies. 

% ======================================================================
\section{Extension of our concept}
\label{sec:extension}

\begin{figure}[!htb]
\centering
\includegraphics[scale=1.00]{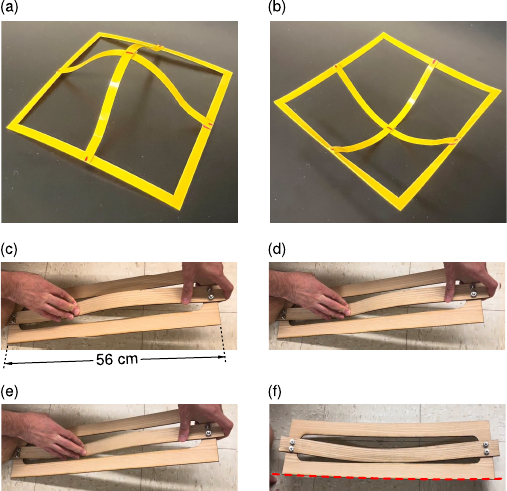}
\caption{Extensions of our structural concept. Bistable 2D unit in its (a) first stable state and (b) second stable state; when tiled in 2D, this unit will produce a true structural analog to bistable dimpled sheets. Up-scaled wooden version of our structural unit (c)-(d) during the snapping process and (f) in its second stable state, with the dashed line highlighting a slight base curvature.}
\label{fig:2d_and_upscaled}
\end{figure}

It must be pointed out that the unit introduced so far is not a true analog of the unit cell of a 2D dimpled sheet. In fact, units of 2D dimpled sheets are strongly coupled -- an aspect that is responsible for the system's ``memory'' -- while our work on forward kinematics and inverse design relies on the absence of such coupling in our structures. It is then legitimate to ask how our unit could be altered to display such coupling when tiled in 2D. Our idea for such structural unit is shown in Fig.~\ref{fig:2d_and_upscaled}a in its undeformed state, and it features a square base and two arches connecting opposite midpoints of the squares' boundaries; \textcolor{black}{this unit presents some resemblance with other dome-shaped structures featuring networks of beams and strips~\cite{baek2018form,poincloux2023indentation}, with the additional feature of being self-standing and not requiring special boundary conditions to be bistable}. 
We can see that snapping the structure to its second stable state, shown in Fig.~\ref{fig:2d_and_upscaled}b, causes a change of curvature of all boundaries. Analyzing the behavior of this unit, how it can be tiled in 2D arrays, or exploring the effect of different design parameters such as the number of edges of the base are all aspects that deserve a separate treatment.

Throughout this article, we also hint at the up-scalability of our structures. To demonstrate that our concept is truly up-scalable, we briefly report on the realization of a larger wooden unit cell that presents some of the attributes of its polymeric, smaller-scale analog. The $\approx 50\,$cm unit, entirely made of $1/8\,$in-thick white ash wood, is shown in Fig.~\ref{fig:2d_and_upscaled}c-e during its snapping process, and in Fig.~\ref{fig:2d_and_upscaled}f in its second stable state. Our fabrication process involves laser cutting of the base and arch, laser cutting of a sinusoidal mold, forming of the wooden arch via steam bending using a custom-built steam box, and assembly of arch and base via nuts and bolts. While this unit presents bistable attributes, the base does not significantly curve in the second state, highlighting some of the challenges in upscaling. First of all, to avoid rupture during snapping, we had to change the inner shape of the base to avoid stress concentrations at the rectangle's corners. Additionally, the lack of overall curving of the base could be due to the orthotropic nature of wood, and to the fact that manufacturing constraints impose that wood fibers be oriented with the longitudinal direction of our unit. This causes the wood to be particularly weak to lateral bending. In turn, when we snap the unit, deformation concentrates near the attachment point, rather than causing an overall bending of the base. These challenges, and the fact that stress and deformation concentrations cannot be predicted by our planar DER model, hint at the need of a more high-fidelity model to simulate these larger units made of structural materials -- something that we also leave for future work.

% ======================================================================
\section{Conclusions}
\label{sec:conclusion}
% ======================================================================

% SUMMARY
In this work, we present a concept of shape retaining morphing structure based on locally-snapping, bistable unit cells. We introduce a fundamental ``arch and base" unit displaying some of the key mechanics of dimpled sheets, illustrate a simple fabrication process, evaluate the unit's load-deformation response through experiments, and compare the results to those of a reduced-order numerical model based on the efficient DER formalism. Then, we explore the parametric space of the fundamental unit and show that the shape of the unit in its second stable state does not significantly depend on the thickness and width of the sheet used to make the structure (for the considered range of parameters). We then focus on the behavior of 1D arrays of units, and predict their shapes by concatenating results of single-unit simulations. We also show an inverse design method to automatically construct arrays of units that, once snapped,  match desired target shapes. Finally, we briefly show how our structural concept could be up-scaled and introduce a unit that, when tiled in 2D, would represent a true structural analog to dimpled sheets. \textcolor{black}{Pending careful studies on the load bearing capacity of our snapped arrays, these could potentially find applications as morphing lightweight architectural structures or as cores for morphing wings.}

Our results suggest that there a few things to improve in future work. First and foremost, to overcome fabrication inconsistencies, we could fabricate our structures out of fiber-reinforced composites -- which can be directly cured into desired, stress-free shapes, thus removing the need for localized, inherently-nonuniform heating. While our DER model does an excellent job at capturing the mechanics and kinematics of our units given its reduced-order nature, it does not allow to accurately predict the response of units with geometrical parameters near the limits of our design space. This could be overcome by extending the DER framework to be able to explicitly model the base as an assembly of rigidly-connected rods. In that case, the model would also have to include torsion and \textcolor{black}{perhaps even} ribbon kinematics~\cite{huang2022discrete}. Such a model could also allow us to simulate the behavior of the 2D unit of Fig.~\ref{fig:2d_and_upscaled}a. Additionally, to more accurately predict the bistability loss observed for some polymeric samples, viscoelasticity should be added to the model.

% ======================================================================
\section*{Acknowledgments}
% ======================================================================

We acknowledge support from the Research Foundation for the State University of New York and from the National Science Foundation (CMMI-2045191). P.C. and A.R. thank Connor McMahan for his help with DER early on in this project.

% ======================================================================
\section*{Author Contribution}
% ======================================================================

A.R.: Methodology; Software; Formal Analysis; Investigation; Resources; Visualization; Writing - Original Draft. S.F.: Conceptualization; Resources. S.T.: Resources; Investigation. C.Z.: Resources; Investigation. P.C.: Conceptualization; Investigation; Resources; Writing - Review \& Editing; Supervision; Funding Acquisition.

% \bibliographystyle{elsarticle-num}

% \bibliography{mybibfile}

% ======================================================================
\appendix
% ======================================================================

% ----------------------------------------------------------------------
\section{Basics on DER and our modifications}
\label{app:der}

In this section, we report some basics on DER for the interested reader. While this account is largely based on the book by Jawed et al.~\cite{jawed2018primer}, we also discuss key differences between our DER implementation and classical DER formulations.

\begin{figure}[!htb]
\centering
\includegraphics[scale=1.00]{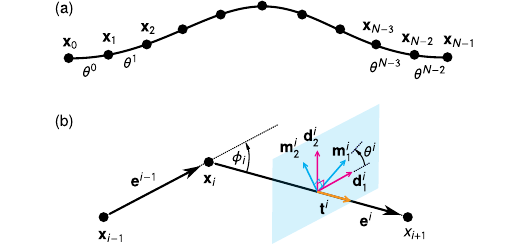}
\caption{(a) A polygonal chain of a discrete elastic rod, consisting of nodes having positional degrees of freedom $\mathbf{x}_0, \mathbf{x}_1, \cdots, \mathbf{x}_{N-1}$, and edges having rotational degrees of freedom $\mathbf{\theta}^0, \mathbf{\theta}^1, \cdots, \mathbf{\theta}^{N-1}$. (b) Each edge vector $\mathbf{e}^i$ is associated with two orthonormal frames: a material frame $\{ \mathbf{m}_1^i, \mathbf{m}_2^i, \mathbf{t}^i \}$ and a reference frame $\{ \mathbf{d}_1^i, \mathbf{d}_2^i, \mathbf{t}^i \}$ to help capture the twist $\theta^i$ in that corresponding edge.}
\label{fig:der_chain}
\end{figure}

\figurename \ref{fig:der_chain}a shows the centerline of a rod which is discretized into $N$ nodes: $\mathbf{x}_0, \mathbf{x}_1, \cdots, \mathbf{x}_{N-1}$, connected by $N-1$ edges: $\mathbf{e}^0, \mathbf{e}^1, \cdots, \mathbf{e}^{N-2}$. 
In the DER literature, node-related parameters are denoted by subscripts, while edge-related quantities are expressed by superscripts. In a 3D problem a rod has $4N-1$ degrees of freedom given by the state vector
\begin{equation}
    \mathbf{q} = 
    \begin{bmatrix}
         \mathbf{x}_0, \theta^0, \mathbf{x}_1, \theta^1, \cdots, \mathbf{x}_i,  \theta^i , \cdots , \mathbf{x}_{N-2},  \theta^{N-2} ,\mathbf{x}_{N-1} 
    \end{bmatrix}^\text{T},
\end{equation}
where $\mathbf{x}_i$ is the position vector and $\theta^i$ is the twist angle. While DER is typically position-based, we alter the formulation to consider displacements as our degrees of freedom, introducing the displacement vector
\begin{equation}
    \mathbf{u} = 
    \begin{bmatrix}
         \mathbf{u}_0, \Delta\theta^0, \mathbf{u}_1, \Delta\theta^1, \cdots, \mathbf{u}_i,  \Delta\theta^i , \cdots , \mathbf{u}_{N-2},  \Delta\theta^{N-2} ,\mathbf{u}_{N-1} 
    \end{bmatrix}^\text{T},
\end{equation}
where $\mathbf{u}_0, \mathbf{u}_1, \cdots, \mathbf{u}_{N-1}$ are nodal displacement vectors and ${\Delta\theta}^0, {\Delta\theta}^1, \cdots, {\Delta\theta}^{N-1}$ are changes in twist angle of the edges.

The edge $\mathbf{e}^{\color{black} i}$ has two orthonormal frames: a material frame $\{ \mathbf{m}_1^i, \mathbf{m}_2^i, \mathbf{t}^i \}$ and a reference frame $\{ \mathbf{d}_1^i, \mathbf{d}_2^i, \mathbf{t}^i \}$ as shown in \figurename \ref{fig:der_chain}b. The two orthonormal frames are separated by the twist angle $\theta^i$, and they share the same unit tangent of the corresponding edge, defined by $\mathbf{t}^i = \mathbf{e}^i / \norm{\mathbf{e}^i}$. The strains and the associated energies in the model are divided into three categories: stretching, bending and twisting, which are summarized in the following.

% ----------------------------------------------------------------------

The stretching strain of an edge is computed as
\begin{equation}
    \epsilon^i = \frac{\norm{\mathbf{e}^i}}{\norm{\bar{\mathbf{e}}^i}} - 1,
\end{equation}
where ``overbar'' quantities refer to the undeformed (reference) state. The stretching energy is written as
\begin{equation}
    E_s = \frac{1}{2} \sum_{i=0}^{N-2} EA^i (\epsilon^i)^2 \norm{\bar{\mathbf{e}}^i}{\color{black},}
\end{equation}
where $E$ is the material's Young modulus and $A^i$ the undeformed cross sectional area of the rod.

% ----------------------------------------------------------------------

The bending strain is computed by the misalignment of the binormals between two adjacent edges at node $\mathbf{x}_i$, given by the discrete integrated curvature vector
\begin{equation}
    (\kappa \mathbf{b})_i = \frac{2\mathbf{t}^{i-1} \times \mathbf{t}^i}{1+\mathbf{t}^{i-1} \cdot \mathbf{t}^i}.
\end{equation}

The norm of this vector, $\norm{(\kappa \mathbf{b})_i} = 2\tan\left(\phi_i/2\right)$, is related to the turning angle $\phi_i$ shown in \figurename \ref{fig:der_chain}b. The material curvatures associated with node $\mathbf{x}_i$ are computed by the inner products of the material frame vectors and the discrete integrated curvature:
\begin{equation}
\begin{aligned}
    \kappa_{k_1} = \frac{1}{2} \left(\mathbf{m}_1^{k-1} + \mathbf{m}_1^{k}\right) \cdot (\kappa \mathbf{b})_k, \\
    \kappa_{k_2} = -\frac{1}{2} \left(\mathbf{m}_2^{k-1} + \mathbf{m}_2^{k}\right) \cdot (\kappa \mathbf{b})_k.
\end{aligned}
\end{equation}
The bending energy is then given by
\begin{equation}
    E_b = \frac{1}{2} \sum_{i=0}^{N-2}  \frac{E{I_i}_1}{\ell_i} ({\kappa_i}_1 - \bar{\kappa_i}_1)^2 + \frac{E{I_i}_2}{\ell_i} ({\kappa_i}_2 - \bar{\kappa_i}_2)^2
\end{equation}
where $\bar{\kappa_i}_1$ and $\bar{\kappa_i}_2$ are the reference curvatures, $\ell_i$ is the Voronoi length (a ``nodal length'' defined as the sum of half the lengths of the edges meeting at a node), and $I_{i1}$, $I_{i2}$ are the second moments of area of the cross section.

% ----------------------------------------------------------------------

Due to the irrelevance of twisting mechanics in our current problem, we refrain from introducing details on twist and twisting energy, and instead point the reader towards the book by Jawed et al.~\cite{jawed2018primer} for more information on the topic.

% The twist at a node $\mathbf{x}_i$ is computed by
% %
% \begin{equation}
%     \tau_i = \theta^i - \theta^{i-1} + m^i_\text{ref},
% \end{equation}
% %
% where $m^i_\text{ref}$ is the reference twist~\cite{jawed2018primer}, a quantity that we . The twisting energy is written as:
% %
% \begin{equation}
%     E_t = \frac{1}{2} \sum_{i=0}^{N-2} GJ_i \frac{(m_i - \bar{m}_i)^2}{\bar{\ell_i}},
% \end{equation}
% %
% where $GJ_i$ is the twisting stiffness of the corresponding edge.

The total energy is given by summing individual energy components:
\begin{equation}
    E = E_s + E_t + E_b.
\end{equation}
The internal forces (and moments) for each positional (or rotational) degree of freedom $u_i$ are expressed by
\begin{equation}
    F^\text{int}_i = \pardydx{}{u_i} (E_s + E_t + E_b).
\end{equation}

% ----------------------------------------------------------------------
\begin{algorithm} [!b]
    \caption{Pseudocode of  our DER algorithm\label{alg:pseudocode}}
    \For{each rod}
    {
        \vspace{0.5 mm}
        Read $N$, cross section height $h$, cross section width $w$, $E$, material density $\rho$, Poisson's ratio $\nu$, $\mathbf{F}^\text{ext}$ and boundary conditions\\

        Compute kinematic properties, i.e., $\mathbf{t}^i$, $\phi_i$, $\mathbf{m}_1^i, \mathbf{m}_2^i$, $\mathbf{d}_1^i, \mathbf{d}_2^i$, twist angle ${\color{black}\theta}^i$, $\kappa_{i1}, \kappa_{i2}$ etc.
    }

    \smallskip
    
    Assemble global $\mathbf{q}$, $\mathbf{F}^\text{ext}$,  $\mathbf{M}$
    
    Set $\mathbf{W}$ (to bind DOFs and incorporate BCs) and solution parameters (choice of algorithms, time step, tolerances, etc.) 
    
    Initialize: $\mathbf{u}=\mathbf{0}$ and velocity vector $\dot{\mathbf{u}}=\mathbf{0}$

    Set time step (increment) counter $j=1$
    
    %\While(\tcp*[f]{\textcolor{blue}{Also functions as time loop}})
    \While{current load $\lambda \mathbf{F}^\mathrm{\color{black}ext}$ $<$ target load $\mathbf{F}^\mathrm{\color{black}ext}$}
    {
        \vspace{0.5 mm} Read latest $\mathbf{u}_j$, $\dot{\mathbf{u}}_j$, $\mathbf{q}_j$ vectors from $\mathbf{u}$, $\dot{\mathbf{u}}$, $\mathbf{q}$\\
        Set $\mathbf{u}_{j+1}=\mathbf{u}_j$, $\mathbf{q}_{j+1}=\mathbf{q}_j$

        \While{residual $>$ tolerance}
        {
            \vspace{0.5 mm}\For{each rod}
            {
                \vspace{0.5 mm}
                Update kinematic parameters \\ %\tcp*[f]{\textcolor{blue}{This is why we need position vectors}}\\
                Compute element $\mathbf{F}^\text{int}$, $\mathbf{K}_t$
            }
            
            \smallskip
            Assemble global $\mathbf{F}^\text{int}$, $\mathbf{K}_t$
            
            \vspace{0.5 mm}\For{each spring}
            {
                \vspace{0.5 mm}Update $\mathbf{F}^\text{int}$ and $\mathbf{K}_t$
            }
            
            Apply $\mathbf{W}$ to get reduced form of $\mathbf{F}^\text{ext}$, $\mathbf{F}^\text{int}$ and $\mathbf{K}_t$
            
            Compute $\mathbf{R} = \mathbf{M} \frac{\mathbf{u}_{j+1} - \mathbf{u}_j}{\Delta t^2} - \mathbf{M}\frac{\dot{\mathbf{u}}_j}{\Delta t} - \mathbf{F}^\text{ext} - \mathbf{F}^\text{int}$ 
            
            Compute $\mathbf{J} = \mathbf{M}\frac{1}{\Delta t^2} + \mathbf{K}_t$
            
            Find new $\mathbf{u}_{j+1}$ and increment in load factor $\Delta\lambda$ using MGDC method
            
           Update $\mathbf{q}_{j+1} = \mathbf{q}_{j} + \mathbf{u}_{j+1}$ and $\dot{\mathbf{u}}_{j+1}=(\mathbf{u}_{j+1}-\mathbf{u}_{j})/\Delta t$ and $\lambda=\lambda+\Delta\lambda$
        }   
        Store $\mathbf{u}_{j+1}$, $\dot{\mathbf{u}}_{j+1}$, $\mathbf{q}_{j+1}$ in $\mathbf{u}$, $\dot{\mathbf{u}}$, $\mathbf{q}$

        Update time step (increment) counter $j=j+1$
    }
\end{algorithm}

The DER formulation is typically dynamic, and we preserve this aspect in our codes, even though our time step $\Delta t$ is chosen to be very large, the external force increases in a ramp-like fashion (by setting the force at each time step to be a fraction $\lambda \mathbf{F}^\text{ext}$ of the target load $\mathbf{F}^\text{ext}$, where $\lambda$ is a load factor), \textcolor{black}{and damping is set to zero,} effectively making our procedure quasi-static. Using a simple forward-Euler time stepping algorithm, the dynamic equation of motion is given by
%
% \begin{equation}
%     \mathbf{M}\mathbf{\ddot{q}} = \mathbf{F^\text{ext}} + \mathbf{F^\text{int}}
% \end{equation}
%
\begin{equation}
    \mathbf{M} \frac{\mathbf{u}_{j+1} - \mathbf{u}_j}{(\Delta t)^2} - \mathbf{M}\frac{\dot{\mathbf{u}}_j}{\Delta t} - \mathbf{F}^{\text{ext}} (t_{j+1}, \mathbf{u}_j, \dot{\mathbf{u}}_j) - \mathbf{F}^{\text{int}} (\mathbf{u}_{j+1}, \dot{\mathbf{u}}_{j+1}) = \mathbf{0},
\end{equation}
where $\mathbf{M}$ is the lumped mass matrix, $\dot{\mathbf{u}}$ is the velocity vector, $\mathbf{F}^\text{ext}$ is the external force vector, and $\mathbf{F}^\text{int}$ is the internal force vector. Subscript $j$ refers to the previous time instant in which all quantities are known, while $j+1$ indicates the quantities we are solving for. The left-hand side of this equation is the residual, labeled $\mathbf{R}$. This nonlinear equation is solved at each time step via a numerical solution technique. While classical DER uses a Newton-Raphson algorithm, we implemented a modified generalized displacement control (MGDC) method~\cite{liu2017nonlinear} to better capture the negative-stiffness parts of the force-displacement curve. During the solution process, we compute the Jacobian matrix $\mathbf{J}$, obtained by taking derivatives of the residual with respect to the unknown $\mathbf{u}_{j+1}$. In the forward-Euler case, the Jacobian is computed as
\begin{equation}
    \mathbf{J} = \mathbf{M}/\Delta t^2 + \mathbf{K}_t,
\end{equation}
where $\mathbf{K}_t$ is the tangent stiffness matrix. Our preference for dealing with tangent stiffness matrices over the Hessians used in classical DER is what pushed us towards a displacement-based formulation.

The pseudocode of our DER formulation is given in Algorithm \ref{alg:pseudocode}. With respect to Jawed's DER algorithm, we also add a constraint matrix $\mathbf{W}$ to account for the boundary conditions and to constrain the relative position of nodes (as we do at the arch location where the we attach the actuating rod). {\color{black}To give a brief account on how the constraint matrix works, we consider a generic DER system whose degrees of freedom (DOFs) are listed in
\begin{equation*}
    \mathbf{u} = 
    \begin{bmatrix}
         \mathbf{u}_0, \Delta\theta^0, \cdots, \mathbf{u}_{j-1},  \Delta\theta^{j-1} ,\mathbf{u}_{j},  \Delta\theta^{j}, \mathbf{u}_{j+1},  \Delta\theta^{j+1} , \cdots , \mathbf{u}_{N-1} 
    \end{bmatrix}^\text{T}.
\end{equation*}
Our intent is to constrain together the displacements in all directions of nodes $j-1$, $j$ and $j+1$, while not constraining the torsional DOFs. We call 
\begin{equation*}
    \mathbf{u}_i = 
    \begin{bmatrix}
         \mathbf{u}_0, \Delta\theta^0, \cdots, \mathbf{u}_{j}, \Delta\theta^{j-1}, \Delta\theta^{j}, \Delta\theta^{j+1}, \cdots , \mathbf{u}_{N-1} 
    \end{bmatrix}^\text{T}
\end{equation*}
the vector of independent DOFs, while $\mathbf{u}_{j-1}$ and $\mathbf{u}_{j+1}$ are dependent DOFs and are constrained to behave as $\mathbf{u}_j$. The constraint matrix $\mathbf{W}$ allows to write $\mathbf{u}$ as a function of $\mathbf{u}_i$, as follows:
\[
    \mathbf{u} = \mathbf{W} \mathbf{u}_i
\]
\[
    \begin{bmatrix}
        \mathbf{u}_0 \\ \Delta\theta^0 \\ \vdots \\ \mathbf{u}_{j-1} \\  \Delta\theta^{j-1} \\\mathbf{u}_{j} \\  \Delta\theta^{j} \\ \mathbf{u}_{j+1} \\  \Delta\theta^{j+1} \\ \vdots \\ \mathbf{u}_{N-1} \\
    \end{bmatrix}
    =
    \begin{bmatrix}
        \mathbf{I} & 0 & \cdots & 0 & 0 & 0 & 0 & \cdots & 0 \\
        0 & 1 & \cdots & 0 & 0 & 0 & 0 & \cdots & 0 \\
        \vdots & \vdots & \ddots & \vdots & \vdots & \vdots & \vdots & \ddots & \vdots \\
        0 & 0 & \cdots & 0 & \mathbf{I} & 0 & 0 & \cdots & 0 \\
        0 & 0 & \cdots & 1 & 0 & 0 & 0 & \cdots & 0 \\
        0 & 0 & \cdots & 0 & \mathbf{I} & 0 & 0 & \cdots & 0 \\
        0 & 0 & \cdots & 0 & 0 & 1 & 0 & \cdots & 0 \\
        0 & 0 & \cdots & 0 & \mathbf{I} & 0 & 0 & \cdots & 0 \\
        0 & 0 & \cdots & 0 & 0 & 0 & 1 & \cdots & 0 \\
        \vdots & \vdots & \ddots & \vdots & \vdots & \vdots & \vdots & \ddots & \vdots \\
        0 & 0 & \cdots & 0 & 0 & 0 & 0 & \cdots & \mathbf{I} \\
    \end{bmatrix}
    \begin{bmatrix}
        \mathbf{u}_0 \\ \Delta\theta^0 \\ \vdots \\ \Delta\theta^{j-1} \\\mathbf{u}_{j} \\  \Delta\theta^{j} \\ \Delta\theta^{j+1} \\ \vdots \\ \mathbf{u}_{N-1} \\
    \end{bmatrix},
\]
where $\mathbf{I}$ is a $3\times 3$ identity matrix. In our code's solution loop, we solve for this reduced displacement vector $\mathbf{u}_i$. Thus, the external force vector, internal force vector, mass matrix and the stiffness matrix are transformed to $\mathbf{W}^\text{T} \mathbf{F}^\text{ext}$, $\mathbf{W}^\text{T} \mathbf{F}^\text{int}$, $\mathbf{W}^\text{T} \mathbf{M} \mathbf{W}$ and, $\mathbf{W}^\text{T} \mathbf{K}_t \mathbf{W}$ prior to solving. In the DER model specific to our structures, $\mathbf{u}_j$ represent the center node of the arch, while $\mathbf{u}_{j-1}$ and $\mathbf{u}_{j+1}$ represent the two adjacent nodes that are constrained to the center node.
}

% ----------------------------------------------------------------------

\section{Experimental force-displacement curves} \label{app:forcedef}

\begin{figure*}[!htb]
\centering
\includegraphics[scale=1.00]{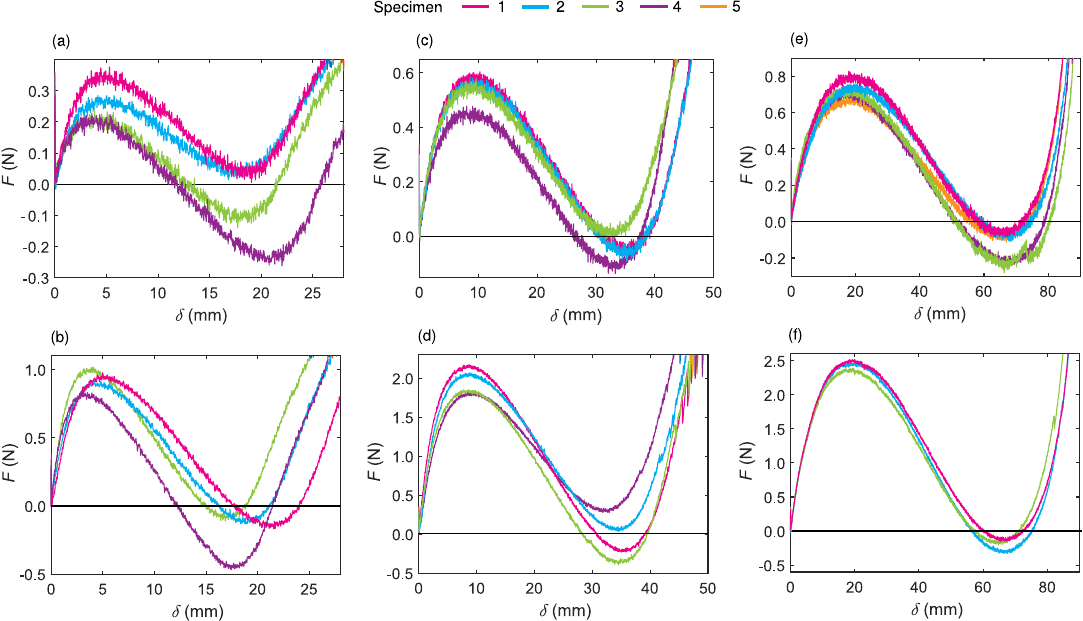}
\caption{Force-deformation responses of all specimens listed in \tablename~\ref{tab:parameters}. (a) $h=10$\,mm, $t=0.5$\,mm; (b) $h=10$\,mm, $t=0.75$\,mm; (c) $h=20$\,mm, $t=0.5$\,mm; (d) $h=20$\,mm, $t=0.75$\,mm; (e) $h=40$\,mm, $t=0.5$\,mm; (f) $h=40$\,mm, $t=0.75$\,mm.}
\label{fig:forcedef_raw}
\end{figure*}

\figurename\ref{fig:forcedef_raw} shows individual force-deformation curves for all the specimens listed in \tablename~\ref{tab:parameters}. The main reason to report these results is to help the reader identify, for each family of design parameters, how many specimens showed bistable behavior and how many did not. 

\figurename\ref{fig:forcedef_raw}a,b show results for arch height $h = 10$ mm, and for thickness $t = 0.5$ mm and $t = 0.75$ mm, respectively. From the plots we can see the variability between the curves is high with respect to other arch heights. The difference between curves is smaller for $t =  0.75$ mm, and all the units are bistable unlike its $t = 0.5$ mm counterpart. Yet, all the curves show large geometrical variability, indicated by the difference of displacement at the zero force points. \figurename\ref{fig:forcedef_raw}c,d show the response for arch height $h = 20$ mm, for $t = 0.5$ mm and $t = 0.75$ mm, respectively. We observe that differences between curves are relatively lower in these plots. However, not all specimens show bistability. One specimen for $t = 0.50$ mm and two specimens for $t = 0.75$ mm were not bistable. Moving to the largest height of $h = 40$ mm, as shown in \figurename\ref{fig:forcedef_raw}e,f for $t = 0.5$ mm and $t = 0.75$ mm, respectively, we observe the most consistent response for both thickness. All the units are bistable and the deviations between specimens are small.

% % ----------------------------------------------------------------------
% \section{Parametric study for asymmetric snapping} \label{a:param}

% \begin{figure*}[!htb]
% \centering
% \includegraphics[scale=1.00]{maps_appendix.pdf}
% \caption{Parametric study maps of normalized heights ($h/L$) vs.\ normalized thicknesses ($t/L$) for critical force for asymmetric snapping} 
% \label{fig:maps_appendix}
% \end{figure*}

% \begin{figure*}[!htb]
% \centering
% \includegraphics[scale=1.00]{matrix_fcr.pdf}
% \caption{Critical forces matrix}
% \label{fig:matrix_fcr}
% \end{figure*}

% \begin{figure*}[!htb]
% \centering
% \includegraphics[scale=1.00]{matrix_deformed.pdf}
% \caption{Deformed shape matrix}
% \label{fig:matrix_deformed}
% \end{figure*}
%

\end{document}